# Light Hadron Masses with 4-GeV Cutoff and L = 2.4 fm

Seyong Kim[a,b,*] and Shigemi Ohta[c]

[a]Center for Theoretical Physics, Seoul National University, Seoul, Korea

[b]Argonne National Laboratory, Argonne, IL60439, USA

[c]Institute of Physical and Chemical Research (RIKEN), Wako-shi, Saitama 351-01, Japan

We discuss preliminary results from our quenched light hadron mass calculation on a $48^3 \times 64$ lattice at the coupling of $\beta = 6.5$. Staggered quarks with masses of $m_q = 0.01, 0.005, 0.0025$ and $0.00125$ are used.

## 1. MOTIVATION

We are calculating light hadron masses in quenched lattice quantum chromodynamics (QCD) with staggered quarks. In a series of works [1–3], one of the authors (SK) was involved in a study of the systematic errors arising from the finite lattice spacing, finite volume, and finite quark mass. An important suggestion from those works is that a combination of the spacelike lattice size of L = 2.4 fm and the coupling of $\beta = 6.5$ should be good enough for extrapolations to remove the errors from finite volume and finite lattice spacing. The errors from finite quark mass is more subtle. Sharpe [4] and Bernard and Golterman [5] suggest that the chiral limit of quenched QCD differs from that of full QCD. In ref. [3], SK and Sinclair argued that indeed the chiral behavior of quenched QCD is different from the full theory in regard of the suggested "quenched chiral log" term. In those works [1–3], however, the three sources of errors were studied separately due to the hardware limitation. In contrast the present work simultaneously addresses the finite-volume and finite-lattice-spacing effects. It uses a VPP500/30 vector-parallel supercomputer at RIKEN [6]. We plan to investigate the chiral limit of quenched QCD in more detail, particularly in regard of a suggestion that the finite volume effect compounds the chiral limit [7]. In the current study the chiral limit should be approached more reliably than in the stronger coupling cases since the flavor symmetry violation is small.

## 2. SIMULATION CHARACTERISTICS

The $48^3 \times 64$ lattice volume and the coupling of $\beta = 6.5$ are selected for the reasons discussed in the above. The spacelike lattice size of 48 makes a 24-node partition of the 30-node VPP500/30 convenient. As was already reported [6], each node of this computer has a peak speed of 1.6 GFLOPS, 256 Mbytes of SRAM memory and an ability to communicate with any of the other nodes at 800 Mbytes/s peak. Through coding in a modified fortran we obtain more than 1 GFLOPS sustained per node for the current work.

In generating the gauge configuration we use a Metropolis update followed by an over-relaxation, as in the earlier works [1–3]. The separation between hadron-mass calculations is 1000 such pairs of Metropolis and over-relaxation sweeps, and take about 3 hours in total including the necessary disk access. Since decorrelation took about 500 such sweeps for the $32^3 \times 64$ lattice at $\beta = 6.5$, the 1000-sweep separation should be good enough. We have not seen any sign that the contrary is true. All the configurations used for the hadron-mass calculations, almost 2 Gbytes each, are stored in a tape archive. This will enable us to study hadrons with strangeness, charm

[*]Talk presented by SK at the International Symposium "Lattice 95" in Melbourne, July 1995. We would like to thank D.K. Sinclair for helpful discussions. We thank the computation center of RIKEN for the use of VPP500/30. SK was supported by Argonne National Laboratory, DOE contract W-31-109-ENG-38 in the beginning of this work and would like to thank RIKEN for inviting him to work on this project. He also appreciates support from Prof.'s H.S. Song and C. Lee of Center for Theoretical Physics at Seoul National University. Prof. S.K. Kim at Seoul National University allowed us to access CERN libraries.

and bottom in the near future.

Before calculating the propagators, we calculate chiral condensate $\langle\bar{\psi}\psi\rangle$ for each masses using bi-linear noise scheme with single noise vector.

The gauge field is fixed to Coulomb gauge by an over-relaxation method prior to quark-propagator calculation. For estimating quark propagators we use the standard conjugate gradient (CG) method using fixed $48^3$-volume corner walls and even point walls as sources and a point sink. Staggered quarks with masses of $m_q = 0.01, 0.005, 0.0025$, and $0.00125$ are used. The physical wall size should be about $(2.4\text{fm})^3$ and the physical quark mass should range from about 5 MeV to 40 MeV since the lattice cutoff is $a^{-1} \simeq 4$ GeV at $\beta = 6.5$ according to the scale set by the $\rho$ meson mass [1]. The convergence criterion for the CG iterations is the norm of the residual vector being equal to or smaller than $1.0 \times 10^{-2}$. It takes from about 1000 (for $m_q = 0.01$) to 8000 (for $m_q = 0.00125$) CG iterations till convergence. The calculations from the chiral condensate and gauge fixing to hadron propagators take about 6 hours in total for each gauge configuration. All the quark propagators, almost 1 Gbytes each, are also stored in the tape archive.

After the hadron propagators are calculated, we fit them using the same procedure as in the previous studies [1–3]: CERN library MINUIT is used for minimization of the correlated $\chi^2$ function and the error bar quoted is for one unit change in $\chi^2$. At the time of the Melbourne meeting we reported the results from 30 gauge configurations. Here we choose to report the results from 50 configurations, an increased statistics in the month following the meeting, and using the full covariance matrix.

## 3. RESULT AND DISCUSSION

In Table 1 and 2 we summarize, respectively, the chiral condensate and light hadron masses for each of the four quark mass values calculated from the 50 gauge configurations. Note that all the results presented here are still preliminary. Complete analysis should wait till we accumulate twice more statistics or 100 configurations. In the tables we also compare the current 50-configuration results on the $48^3 \times 64$ volume with those of the 100-configuration ones on the $32^3 \times 64$ lattice [1]. From this comparison it is clear that the current preliminary results from the $48^3$ volume are consistent with the established results from the $32^3$ one. On the other hand we still do not understand why the current results tend to give smaller error bars despite their lower statistics: perhaps through smaller fluctuations due to the larger lattice volume, or perhaps the $\chi^2$ fitting procedure is under-estimating the error. A different error estimate such as those based on the Jack-knife method would be helpful, and is planned after we accumulate more data.

Table 1
Chiral condensate $\langle\bar{\psi}\psi\rangle$ at $\beta = 6.5$.

| $m_q a$ | $48^3 \times 64$ | $32^3 \times 64$ |
|---------|------------------|------------------|
| 0.01    | 0.04467(3)       | 0.04466(4)       |
| 0.005   | 0.02382(4)       | 0.02387(7)       |
| 0.0025  | 0.01337(6)       | 0.01348(9)       |
| 0.00125 | 0.00808(8)       | -                |

It is also worthy to note that the current pion effective mass, plotted in Figure 1, gives better plateau than that from the previous $32^3$-volume calculations though the noise sets in early in time coordinate due to lower statistics. This suggests that the $(2.4\text{fm})^3$ wall is indeed optimal. Interestingly, in the previous calculations [3] at $\beta = 6.0$, effective mass plot for $24^3 \times 64$ also shows better plateau than that for $32^3 \times 64$ or $16^3 \times 64$.

Figure 2 gives the Edinburgh plot. The errors in the figure are obtained by assuming that the relative error in each quantity is independent of each other. In the plot we used the nucleon mass from the "even-point-wall" source because the quality of fitting is better than that from the "corner-wall" source. The other quantities are from the "corner-wall" source. In contrast to the same plot for the $32^3$ volume, all the points, except for the leftmost point for the lightest quark mass, lie below the guiding lines. This suggests that ground-state nucleon is less squeezed on the $48^3$ volume than on the $32^3$.

At the lightest quark mass of $m_q = 0.00125$, we now see the ratio $m_\pi/m_\rho$ as low as $\sim 0.3$. Since we have $m_\pi L \simeq 2.9$, we can probably conclude that our lattice is big enough even down to such a light quark mass. Thus pushing the quark mass $m_q$ even smaller and reaching the region where $m_\pi/m_\rho \simeq 0.2$ with $m_\pi L \simeq 2.0$ and $m_\pi \simeq 0.04$ seem viable. In other words extrapolation to small $m_q$ in the quenched spectrum calculations may soon become unnecessary.

Table 2
Hadron masses at $\beta = 6.5$ on $48^3 \times 64$ lattice.

| $m_q a$ | particle | fitting range in $t$ | mass | mass from $32^3$ |
|---|---|---|---|---|
|  | $\pi$ | 4-17 | 0.1630(6) | 0.1592(10) |
|  | $\pi_2$ | 6-25 | 0.1605(7) | 0.1611(12) |
|  | $\rho$ | 8-17 | 0.2489(15) | 0.2417(24) |
|  | $\rho_2$ | 7-17 | 0.2483(13) | 0.2416(36) |
| 0.01 | $a_1$ | 7-17 | 0.3489(30) | 0.3481(49) |
|  | $b_1$ | 10-17 | 0.3406(80) | 0.3498(81) |
|  | $\sigma$ | 6-20 | 0.3062(42) | 0.3197(62) |
|  | $N_1$ | 10-18 | 0.3678(25) | 0.3761(34) |
|  | $N_2$ | 3-18 | 0.3444(17) | 0.3522(29) |
|  | $\Delta$ | 7-16 | 0.4141(26) | 0.4193(48) |
|  | $\pi$ | 6-18 | 0.1179(10) | 0.1133(13) |
|  | $\pi_2$ | 6-20 | 0.1172(12) | 0.1162(18) |
|  | $\rho$ | 8-20 | 0.2285(23) | 0.2179(42) |
|  | $\rho_2$ | 7-17 | 0.2312(23) | 0.2208(59) |
| 0.005 | $a_1$ | 5-17 | 0.3261(33) | 0.3296(80) |
|  | $b_1$ | 8-20 | 0.3251(73) | 0.3348(186) |
|  | $\sigma$ | 6-20 | 0.3044(94) | 0.3455(210) |
|  | $N_1$ | 10-20 | 0.3187(44) | 0.3416(58) |
|  | $N_2$ | 12-31 | 0.3013(50) | 0.3106(54) |
|  | $\Delta$ | 5-29 | 0.3989(29) | 0.3934(63) |
|  | $\pi$ | 6-18 | 0.0855(12) | 0.0821(17) |
|  | $\pi_2$ | 6-20 | 0.0884(19) | 0.0893(34) |
|  | $\rho$ | 8-20 | 0.2186(37) | 0.2043(65) |
|  | $\rho_2$ | 4-18 | 0.2317(26) | 0.2049(109) |
| 0.0025 | $a_1$ | 6-28 | 0.3076(38) | 0.3063(67) |
|  | $b_1$ | 5-18 | 0.3228(72) | 0.3323(468) |
|  | $\sigma$ | 6-20 | 0.3233(200) | - |
|  | $N_1$ | 8-19 | 0.3045(80) | 0.3287(115) |
|  | $N_2$ | 11-29 | 0.2614(93) | 0.2859(88) |
|  | $\Delta$ | 4-22 | 0.4007(42) | 0.3683(157) |
|  | $\pi$ | 6-17 | 0.0633(15) | - |
|  | $\pi_2$ | 4-13 | 0.0608(38) | - |
|  | $\rho$ | 4-18 | 0.2214(39) | - |
|  | $\rho_2$ | 4-18 | 0.2338(52) | - |
| 0.00125 | $a_1$ | 4-18 | 0.3028(51) | - |
|  | $b_1$ | 5-18 | 0.3083(134) | - |
|  | $\sigma$ | 3-13 | 0.2735(155) | - |
|  | $N_1$ | 4-20 | 0.3318(82) | - |
|  | $N_2$ | 3-23 | 0.2891(58) | - |
|  | $\Delta$ | 4-22 | 0.4029(52) | - |

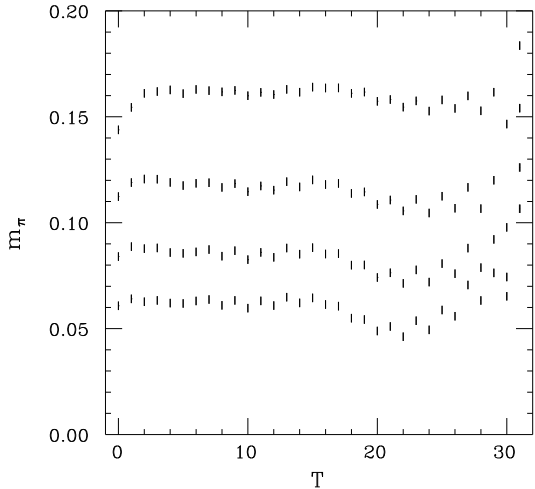

Figure 1. Pion effective mass at $\beta = 6.5$ on $48^3 \times 64$ lattice for quark masses $m_q = 0.01, 0.005, 0.0025$ and $0.00125$. Corner-wall source.

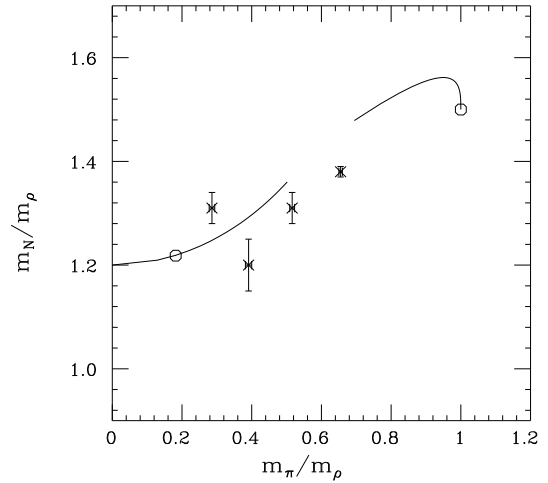

Figure 2. Edinburgh plot at $\beta = 6.5$ on $48^3 \times 64$ lattice for quark masses $m_q = 0.01, 0.005, 0.0025$ and $0.00125$. The two circles are the experimental point (left) and naive quark model prediction (right).

Also the current result shows approximate flavor symmetry restoration: the masses of $\pi$ and $\pi_2$ agree with each other for all the calculated quark masses and the masses of $\rho$ and $\rho_2$ agree for the two heavier quark masses and are close to each other for the lighter two.

All these attractive features of the current results suggest that the chiral limit of our data will be more interesting. However, we think that the investigation on the quenched chiral log in the pion mass $m_\pi$ and chiral condensate $\langle \bar\psi \psi \rangle$ should wait for a still higher statistics of at least 100 gauge configurations. We do not think the current statistics is enough for that purpose because a) each point in the pion effective mass plot shows large fluctuation despite the fact that the error estimated by the full covariance matrix is small, and b) the $\rho$-meson mass data also imply large fluctuation as $m_\rho$ at $m_q = 0.00125$ is similar to, and not smaller than, that at $m_q = 0.0025$.